\documentclass[aps,pre,onecolumn,superscriptaddress,amsmath,amssymb]{revtex4-2}

\usepackage{amsmath}
\usepackage{amssymb}
\usepackage{dsfont} % for unit operator  (indicator function) \mathds{1}

\usepackage{graphicx}
\usepackage[dvipsnames]{xcolor}
\usepackage{color}

\newcommand{\be}{\begin{equation}}
\newcommand{\ee}{\end{equation}}
\newcommand{\bel}[1]{\begin{equation}\label{#1}}
\newcommand{\bea}{\begin{eqnarray}}
\newcommand{\eea}{\end{eqnarray}}
\newcommand{\balign}{\begin{align}}
\newcommand{\ealign}{\end{align}}
\newcommand{\ba}{\begin{array}}
\newcommand{\ea}{\end{array}}
\newcommand{\bfig}{\begin{figure}}
\newcommand{\efig}{\end{figure}}

\newcommand{\eref}[1]{(\ref{#1})}

\allowdisplaybreaks

\newcommand{\exval}[1]{\mbox{$\langle \, {#1}\, \rangle$}}

\newcommand{\rmd}{\mathrm{d}}
\newcommand{\rme}{\mathrm{e}}

\newcommand{\ddt}{\frac{\rmd}{\rmd t}}

\newcommand{\half}{\frac{1}{2}}

\newcommand{\R}{{\mathbb R}}

\newcommand{\Z}{{\mathbb Z}}

\newcommand{\bfeta}{\boldsymbol{\eta}}

\newtheorem{theo}{Theorem}[section]

\newtheorem{df}[theo]{Definition}
\newtheorem{prop}[theo]{Proposition}
\newtheorem{cor}[theo]{Corollary}
\newtheorem{rem}[theo]{Remark}
\newcommand{\proof}{\noindent {\it Proof: }}
\def\qed{\hfill$\Box$\par\medskip\par\relax}

\global\long\def\vfi{\varphi}

\global\long\def\si{\sigma}

\begin{document}

\title{Quest for the golden ratio universality class 
}

\author{Vladislav Popkov}
 \affiliation{Department of Physics,
  University of Wuppertal, Gaussstra\ss e 20, 42119 Wuppertal,
  Germany}
\affiliation{Faculty of Mathematics and Physics, University of Ljubljana, Jadranska 19, SI-1000 Ljubljana, Slovenia}
\author{G.M.~Sch\"utz}
\affiliation{ Departamento de Matemática,
Instituto Superior Técnico,
Av. Rovisco Pais 1, 1049-001 Lisboa, Portugal}

\begin{abstract}

Using mode coupling theory the conditions for all allowed dynamical universality classes 
for the conserved modes in one-dimensional driven systems are 
presented in closed form as a function of the 
stationary currents and their derivatives. With a view on the search for the golden ratio 
universality class the existence of some families of microscopic models is ruled
out a priori by using an Onsager-type macroscopic current symmetry. 
In particular, if the currents are symmetric or antisymmetric under the interchange of the conserved densities, 
then at equal mean densities the golden modes can only appear in the antisymmetric case and 
if the conserved quantities are correlated, but not in the symmetric case where at equal densities one mode 
is always diffusive and the second may be either Kardar-Parisi-Zhang (KPZ), modified KPZ, 3/2-Lévy, or also diffusive.
We also show that the predictions of mode coupling theory for
a noisy chain of harmonic oscillators are exact.

\end{abstract}

\maketitle

\section{Introduction}

The golden ratio $\vfi=(1+ \sqrt{5})/2\approx 1.618$ is undoubtedly one of  the most fascinating irrational numbers. It shows up in completely different circumstances, from in the arrangement of seeds 
in sunflowers or petals in flowers \cite{Livio,1998Levitov}, in the shape of snowflakes \cite{Kepler1611}, the statistics of signals in the human brain \cite{2010Pletzer} to art and architectural masterpieces \cite{2018Meisner}.  From the mathematical viewpoint,  the golden ratio is the 
``most'' irrational number, in the sense that the number for which rational approximations like the 
continued fraction \cite{Khinchin} perform the worst.  

In the realm of physics the golden ratio is somewhat less prominent. In particular, in the theory of phase transitions -- where {\it rational} numbers are ubiquitous in the characteristics of universality classes -- specific irrational numbers have not been encountered until recently.  Indeed,  all   critical indices coming from Landau theory of phase transitions \cite{1937Landau} are rational numbers. Also the critical exponents produced in conformal field theories are all rational (unless continuously varying). 
Generically,  since  various scaling relations relating different critical exponents have the form of a ratio,  the  rationality of critical exponents seems to be a robust feature, leaving little room for irrationality.

It has therefore come as a  surprise that the golden ratio can characterize universal fluctuations
of long-lived relaxation modes of interacting many-body systems with conservation laws,  governed by the dynamical exponent $z$, i.e., with stationary space-time correlations of the form
\begin{align}
\langle \si(0,0) \si(x,t) \rangle \sim f \left( \frac{|x-vt|^z}{t} \right), \label{eq:DynExp}
\end{align}
at  large space and time scales. Here $ \si(x,t)$ is a conserved fluctuation field and $v$ is its characteristic mode velocity. The
theory of the Fibonacci universality classes \cite{Popk15a}, based on mode-coupling theory for nonlinear flucuating hydrodynamics, predicts the possibility of coexistence of modes with different dynamical exponents from the discrete infinite set of Kepler ratios $z_k= C_{k+1}/C_k$ where $C_k$ are the Fibonacci numbers $1,1,2,3,5,\ldots$.  The ubiquitous diffusive Edwards-Wilkinson (EW) mode $k=2$ \cite{Krug97} and the celebrated superdiffusive Kardar-Parisi-Zhang (KPZ) mode $k=3$ \cite{Halp15} appear already in systems with one conservation law,  while to reach the more exotic modes with  $k\geq 4$ at least $k-2$ conservation laws (and hence relaxation modes) are necessary.  The golden ratio 
 $\vfi = \lim_{k\rightarrow \infty} z_k=(1+ \sqrt{5})/2$  turns out to be a remarkable exception: This truly irrational dynamical exponent
can be generated in (\ref{eq:DynExp}) already in a system with just two long-lived relaxation modes. 

With this knowledge in hand, and using fine tuning of interaction parameters we were able to demonstrate the 
existence of the irrational critical exponent $\vfi$ in driven lattice gases with two and three conservation laws \cite{Popk15a,Popk15b}. Also in anharmonic chains with two conservation laws the golden universality class appears for fine-tuned parameter values \cite{Spoh15}.
However,  ideally, one would like to have the means to produce a system from the remarkable golden universality class in a robust way, based on symmetries, instead of fine-tuning. Our present work presents a systematic study of properties of the interactions
needed to generate one or another universality class, including the golden ratio, for a system with two conservation laws. In order to emphasize that under the given assumptions some of the results are mathematically rigorous, we use a mathematical presentation in terms of propositions, theorems and corollaries when appropriate.

\section{Fluctuations in systems with local conservation laws}

To set the stage we describe the setting that we have in mind. In this introductory section 
we keep the number of conserved quantities arbitrary (but finite) and only point to the case of two conservation laws in some instances to facilitate the detailed discussion of two conservation laws in the subsequent sections.

\subsection{Nonequilibrium steady state (NESS)}

Consider a one-dimensional many-body dynamics with translation invariant short-range interactions with $n$ locally conserved densities $\rho_{\alpha}$, $\alpha\in\{1,\dots,n\}$ where a driving force produces macroscopic stationary currents denoted by $j^\alpha$. The resulting non-equilibrium steady state (NESS) is assumed to be unique for given values of the conserved densities,
to have non-vanishing global fluctuations of the conserved densities,
and exhibit a decay of local density correlations that is sufficiently fast to guarantee that the total stationary density fluctuations are finite. These properties 
are generic since in one space dimension stationary correlations
between local observables usually decay rapidly with distance \cite{Spoh91}.
Indeed, for a single conservation law and translation invariant short-range interactions no counter examples to these properties are known except for some very special cases, viz., in systems with long-range interactions \cite{Spoh99,Popk11,Kare17} or with facilitated dynamics \cite{Helb99,Ross00,Gabe10,Blon20,Erig23}.  For two conservation laws long-range correlations accompanied by phase separation have been reported and analysed numerically and by mean field theories \cite{Rama02,Kafr03,Chak16}. All these atypical types of behaviour as well as frozen systems without
density fluctuations are ruled out in the present investigation.

The total stationary density fluctuations are encoded in the compressibility matrix $K$
in terms of the variances
$K_{\alpha\alpha}$ of the conserved quantities and their covariances $K_{\alpha\beta}$. Hence $K$ is symmetric and all matrix elements are in general functions of all the conserved quantities $\rho_{\alpha}$.
The cross correlations given by the covariances may
be positive or negative or even vanish for all pairs of densities while
the variances $K_{\alpha\alpha}$ strictly positive real numbers 
(excluding the atypical situations mentioned above). 
Specifically, for two conservation laws we denote the matrix elements of the compressibility matrix $K$ by
\bel{Kdef}
K = \left( \ba{cc} \kappa_1 & \bar{\kappa} \\ \bar{\kappa} & \kappa_2 \ea \right) .
\ee
rather than by $K_{\alpha\beta}$ and we usually omit the dependence of these quantities
on $\rho_1$ and $\rho_2$.

Also the stationary currents $j^\alpha$ are in general functions of  all 
densities $\rho_{\alpha}$. Their derivatives w.r.t. the densities are denoted
by subscripts $j^\alpha_{\beta}$, $j^\alpha_{\beta\gamma}$, and so on. 
The current Jacobian $J$ has matrix elements $J_{\alpha\beta} =
j^\alpha_{\beta}$ and the Hessians $H^\alpha$ associated with the current-density relation have matrix elements
$H^\alpha_{\beta\gamma} = j^\alpha_{\beta\gamma}$.
Thus for two conservation laws
the current Jacobian and the Hessians are the matrices
\bel{JHdef}
J = \left( \ba{cc} j^1_{1} & j^1_{2} \\ j^2_{1} & j^2_{2} \ea \right), \quad
%\left( \ba{cc} J_{11} & J_{12} \\ J_{21} & J_{22} \ea \right)
%= \left( \ba{cc} (1-2\rho_1)(a_1+b \rho_2) & b \rho_1 (1-\rho_1) \\ b \rho_2 (1-\rho_2) & (1-2\rho_2)(a_2+b \rho_1) \ea \right)
H^\alpha = \left( \ba{cc} j^\alpha_{11} & j^\alpha_{12} \\ j^\alpha_{21} & j^\alpha_{22} \ea \right) 
\ee
with the dependence on $\rho_1$ and $\rho_2$ also usually omitted.

\subsection{Large scale dynamics}

The long-time relaxation of the system is assumed to be determined by
the long-wave length Fourier 
modes of the conserved quantities in the sense that at sufficiently large times
the system is locally (on a microscopic scale) in a stationary state with densities
$\rho_\alpha$, which vary only slowly with space and time. 

The generic decay of correlations ensures bounded density fluctuations 
in the NESS and thus in driven nonequilibrium systems with non-vanishing stationary currents the validity of a coarse-grained
hydrodynamic description of the conserved densities $\rho_\alpha(x,t)$  in terms of a 
continuity equation 
$\partial_t \rho_\alpha(x,t) + \partial_x J_\alpha(x,t) = 0$
where the currents $J_\alpha(x,t)$ depend on macroscopic space and time only via the densities $\rho_1(x,t),\dots,\rho_n(x,t)$ \cite{Spoh91,Kipn99}.
Thus the hydrodynamic limit is given by
a system of hyperbolic conservation laws
\bea 
\partial_t \rho_\alpha(x,t) + \sum_{\beta=1}^{n} 
J_{\alpha\beta} \partial_x \rho_\beta(x,t) = 0
\label{hscl}
\eea
where 
%the $\rho_\alpha(x,t) = \exval{\eta_\alpha(x,t)}$ is the stationary expectation of a 
%microscopic locally conserved quantity $\eta_\alpha(x,t)$ and 
the current Jacobian $J_{\alpha\beta} = \partial_{\rho_\beta} j^\alpha(\rho_1,\dots,\rho_n)$ is given by the stationary current-density relation 
$j^\alpha(\rho_1,\dots,\rho_n)$ evaluated at the space-time point $(x,t)$.

Fluctuations are not captured in this deterministic large-scale description. They are expected to be described by the theory of nonlinear fluctuating hydrodynamics \cite{Spoh14,Spoh16} for fluctuation fields 
which are centered around a fixed stationary mean $\rho_\alpha$ of the conserved quantities.
To analyze the large-scale behaviour of these fluctuations it is most
convenient to express them as the eigenmodes $\sigma_\alpha(x,t)$
that appear in generic form in \eref{eq:DynExp} without mode index $\alpha$,
i.e., as those linear combinations of the conserved fields for which the
current Jacobian $J$
is diagonal for the chosen stationary densities $\rho_\alpha$. According to nonlinear fluctuating hydrodynamics
these eigenmodes  then satisfy the system of coupled stochastic Burgers equations 
\be 
\partial_t \sigma_\alpha + \partial_x  \left(v_\alpha \sigma_\alpha
+ \sum_{\beta,\gamma=1}^{2} G^\alpha_{\beta\gamma} \sigma_\beta \sigma_\gamma
- \sum_{\beta=1}^{2} 
D_{\alpha\beta} \partial_x \sigma_\beta + \xi_\alpha \right) = 0
\label{CSBE}
\ee
with the mode velocities $v_\alpha$ which are the
eigenvalues of the current Jacobian, 
with phenomenological diffusion matrix $D$ and with Gaussian white noises $\xi_\alpha$.
The strength
of the nonlinearity is given by the mode coupling coefficients $G^\alpha_{\beta\gamma}$
which depend primarily on the current Jacobian and also -- in an insubstantial way -- on the static compressibilites $K_{\alpha\beta}$
(see below).

These nonlinear stochoastic partial differential equations (spde's) can be treated by
mode coupling theory as described in \cite{Spoh14}. With this approach we found in \cite{Popk15a,Popk16} together with our collaborators the exact analytical scaling solution for the dynamical structure function \eref{eq:DynExp} for all modes $\alpha$. This yields the Fibonacci dynamical universality classes as follows:
\begin{itemize}
\item In the absence of the diagonal mode coupling term, i.e., 
$G^\alpha_{\beta\beta} = 0$ for all modes $\beta$ 
(including the mode $\beta=\alpha$) the mode $\alpha$ is diffusive and belongs to the Edwards-Wilkinson universality class
with dynamical exponent $z_\alpha=2$.\\
\item For non-vanishing self-coupling term, i.e., 
$G^\alpha_{\alpha\alpha} \neq 0$, 
the mode $\alpha$ is superdiffusive and belongs to the Kardar-Parisi-Zhang universality class
or a modification thereof, both with dynamical exponent $z_\alpha=3/2$.\\
\item If $G^\alpha_{\alpha\alpha} = 0$ but $G^\alpha_{\beta\beta} \neq 0$ for some
mode $\beta\neq\alpha$ then mode $\alpha$ is in a subdiffusive L\'evy
universality class with the dynamical exponent $z_\alpha$ being a Kepler ratio of two consecutive Fibonacci numbers or the golden ratio.
\item The non-diagonal mode coupling terms $G^\alpha_{\beta\gamma}$
with $\beta\neq\gamma$ are irrelevant for these scaling properties.
\end{itemize}
The results of \cite{Popk15a,Popk16} for the scaling functions are valid for strict hyperbolicity of
the underlying deterministic pde \eref{hscl}, i.e, when the
eigenvalues of the current Jacobian are non-degenerate.

It is worth stressing that the mode coupling coefficients $G^\alpha_{\beta\beta}$ appearing in the macroscopic phenomenological spde \eref{hscl} are fully given by the stationary distribution of the underlying dynamics, viz.,  by the current Jacobian $J$ and by the static compressibility matrix $K$. In fact, whether $G^\alpha_{\beta\beta}$ vanishes or not is determined by the current Jacobian $J$ alone, the compressibilities encoded in $K$ only renormalize the amplitudes. This implies the 
very remarkable fact that  all the {\it dynamical} universality classes of a system can be
read off from the {\it stationary} currents alone.  An explicit example is provided in the Appendix~\ref{app:NLIEforOscillators}.

Specifically for two conservation laws, treated independently in \cite{Popk15b} and \cite{Spoh15} 
in the same issue of J. Stat. Phys., the elusive golden ratio 
occurs if and only if $G^1_{11} = G^2_{22} = 0$
and $G^1_{22} \neq 0$ and $ G^2_{11} \neq 0$. In this case both modes belong to
the golden universality class with $z_1=z_2=\varphi$. 
However, the question that has been left open and which is addressed here are the circumstances under which these conditions may be satisfied or not, i.e., which current-density relations 
that arise from some microscopic dynamics admit or forbid the occurrence of two golden modes.

\subsection{Onsager-type current symmetry and microscopic dynamics}

Before attacking this problem directly we point out a
further general consequence of the generic assumption of sufficiently rapidly
decaying stationary correlations. This is the Onsager-type current symmetry 
\bel{currsym}
JK = KJ^T
\ee 
proved rigorously first for a family of Markovian particle systems with fully factorized 
stationary product measure in \cite{Toth03} and later generally under very mild assumptions concerning
the decay of stationary correlations and range of microscopic interactions for classical dynamics in \cite{Gris11} and subsequently for quantum systems in \cite{Cast16,Kare19}.
This current symmetry appears in many contexts in hydrodynamic theory, 
see e.g. \cite{Spoh14,Spoh16} for a review and \cite{Doyo17,Alba21,DeNa22}
for recent applications in generalized hydrodynamics for integrable quantum systems. As pointed out in \cite{Toth03} the symmetry \eref{currsym}
can be seen as a non-equilibrium version of the Onsager reciprocity relations in so far as its validity relies only on general time-reversal properties
of the time evolution and on an extremely mild assumption on the decay of correlations as discussed in the rigorous proofs of \cite{Gris11,Kare19}. Indeed, the proof of \eref{currsym} follows arguments analogous to those for the Onsager relations.
We shall call one-dimensional physical systems with these generic properties
{\it regular one-dimensional hydrodynamic systems}. 
%As a consequence of the current symmetry \eref{currsym}, the eigenvalues of the current Jacobian are real as it has to be on physical grounds since they are the mode velocities.

Remarkably and perhaps also surprisingly, this 
{\it macroscopic} current symmetry directly imposes constraints on the existence
of specific {\it microscopic} dynamics: For a given family of 
invariant measures (parametrized by the conserved densities
$\rho_\alpha$ which fix $K$) the only physically permissible microscopic dynamics are such 
that the transition rates yield currents (which fix $J$) compatible with 
\eref{currsym}. 

To appreciate the significance of this constraint in the search for physical
systems that exhibit the golden
universality class (or any other specific Fibonacci universality class) one must recall that
for predicting universality classes for some microscopic dynamics
one needs to know which diagonal mode coupling
matrix elements $G^{\alpha}_{\beta\beta}$ vanish. This requires knowledge of the
{\it exact} stationary current-density relation, which in turn requires knowledge
of the exact invariant measure of the dynamics. However, away from
thermal equilibrium detailed balance does not hold, and there is no simple
recipe for obtaining the exact invariant measure for a given dynamics.
Therefore, computing the invariant invariant for some ``nice'' physical
dynamics is usually a hopeless endeavor.

Instead, in the search for specific universality classes one usually 
works in opposite direction:\\ (i) One starts
by defining a measure with a simple structure (such as product or matrix product
measures \cite{Blyt07,Cant22}) that allows for computing expectation values like the stationary currents
in explicit form as functions of the densities,\\ (ii) proposes a physically 
motivated dynamics (such as short-range interactions) that respects the conservation laws,\\
(iii) proves invariance of the measure w.r.t. this dynamics and/or adjusts
parameters accordingly,\\ (iv) computes stationary currents and compressibilities 
from (i) and (ii).\\
However, there is no obvious starting point for guessing such a hypothetical 
dynamics (which might not even exist) 
and then checking invariance w.r.t. the measure
may be cumbersome due to the absence of detailed balance. 
Therefore, being able to rule out {\it a priori} certain microscopic dynamics as unphysical for a given invariant measure can be useful in the search for dynamical universality classes.

Here the current symmetry helps. It does not provide a sufficient condition
for a microscopic dynamics to have the specified measure as invariant measure, but a necessary condition that can be used to rule out a proposed dynamics. 
This property can be utilized by taking step (iv) {\it before}
step (iii): If step (iv) yields a current-density relation 
that is incompatible with \eref{currsym} then there is no need to go through
the potentially daunting task of step (iii).
Unlike step (iii) (which may actually be an unsoluble problem), step (iv) is 
easy
since the measure has been proposed in step (i) so as
to facilitate the computation of the stationary currents. 

To illustrate the point we focus on two conservation laws and note that the current symmetry \eref{currsym} then becomes
\bel{currsym2}
J_{21} \kappa_1 - J_{12} \kappa_2 = ( J_{11} - J_{22}) \bar{\kappa} .
\ee
Evidently, this relation constitutes a necessary condition 
for the existence of microscopic
dynamics that would be compatible with a predefined invariant measure 
-- without the need to perform an explicit check of invariance of the measure under the proposed dynamics. Only the static compressibilities and the stationary currents
of the model need to be computed using the invariant measure.

This is highlighted in the following theorem
for partially factorized invariant measures
of the form $\pi_{\rho_1,\rho_2} = \pi^1_{\rho_1} \pi^2_{\rho_2}$
where $\pi^\alpha_{\rho_\alpha}$ are measures that do not necessarily
factorize further such as matrix product measures. 
Frequently studied models of this type are two-lane systems 
\cite{Gros03,Popk04,Schm21} where particles of species $\alpha$ 
jump on along a one-dimensional lattice called lane $\alpha$ like in a system
with only one conservation law, but with a rate that
depends on the particle configuration on the other lane.

\begin{theo}
\label{Theo:currsym2}
Any regular hydrodynamic two-component system with a partially 
factorized invariant measure 
\bel{pfim}
\pi_{\rho_1,\rho_2} = \pi^1_{\rho_1} \pi^2_{\rho_2}
\ee
has stationary currents of the form
\bea
j^1(\rho_1,\rho_2) & = & f_1(\rho_1) + \kappa_1(\rho_1) 
\int^{\rho_2} \rmd y \, g(\rho_1,y), \label{j1fac} \\
j^2(\rho_1,\rho_2) & = & f_2(\rho_2) + \kappa_2(\rho_2) \int^{\rho_1} \rmd x \, g(x,\rho_2),  
\label{j2fac}
\eea 
 with the compressibilities $\kappa_\alpha(x) \in \R^+$ for $ x \in \R$, 
functions $f_\alpha: \R \to \R$ and interaction kernel
$g(x,y) \in \R$ for $(x,y) \in \R^2$ that is the same for both currents.
Moreover,  for any pair of densities $(\rho_1,\rho_2)$ the off-diagonal elements of the 
current Jacobian either both vanish or have equal sign.
\end{theo}

\proof The lower integration limits in \eref{j1fac} and \eref{j2fac}
do not need to be specified as they can be absorbed into the functions $f_\alpha(\cdot)$.
By writing the currents in a generic fashion as
\bea
j^1(\rho_1,\rho_2) & = & f_1(\rho_1) + \kappa_1(\rho_1) 
\int^{\rho_2} \rmd y \, g_1(\rho_1,y) \nonumber \\
j^2(\rho_1,\rho_2) & = & f_2(\rho_2) + \kappa_2(\rho_2) \int^{\rho_1} \rmd x \, g_2(x,\rho_2)  \nonumber
\eea 
in terms of functions $f_\alpha(\cdot)$, $g_\alpha(\cdot,\cdot)$, yields
\be 
J_{12}(\rho_1,\rho_2) = \kappa_1(\rho_1) g_1(\rho_1,\rho_2), \quad J_{21}(\rho_1,\rho_2) = \kappa_2(\rho_2) g_2(\rho_1,\rho_2).
\nonumber
\ee 
For a factorized invariant measure one has
$\bar{\kappa} = \partial_{1} \kappa_2 =
\partial_{2} \kappa_1 = 0$ so that the current symmetry reduces to
\bel{currsymfac}
J_{21}(\rho_1,\rho_2) \kappa_1(\rho_1) = J_{12}(\rho_1,\rho_2) \kappa_2(\rho_2).
\ee
Ssince $\kappa_\alpha(\cdot) > 0$ by assumption,  this implies $g_2(\rho_1,\rho_2) = g_1(\rho_1,\rho_2)$ for all $(\rho_1,\rho_2) \in \R^2$ and proves also the assertion regarding the off-diagonal elements of the 
current Jacobian. \qed

For a monotone increase or decrease of the current $j^\alpha$ as 
a function of the other density $\rho_\beta$ with $\beta \neq \alpha$ 
as in the models of
\cite{Gros03,Popk04,Schm21}
and also assuming the existence of symmetries between the densities 
the current symmetry \ref{Theo:currsym2}
provides the following rather general no-go corollary.

\begin{cor}
A regular one-dimensional hydrodynamic two-component system with the three properties\\
(i) partially factorized invariant measure 
$\pi_{\rho_1,\rho_2} = \pi^1_{\rho_1} \pi^2_{\rho_2}$\\
(ii) antisymmetric currents $j^2(\rho_1,\rho_2) = - j^1(\rho_2,\rho_1)$\\
(iii) monotonicity $j^1_2 > 0$ and $j^2_1 > 0$
(or $j^1_2 < 0$ and $j^2_1 < 0$) for all densities $\rho_1,\rho_2$\\
does not exist.
\end{cor}

This holds since on the one hand monotonicity for all $(\rho_1,\rho_2)$ requires that 
$g(\cdot,\cdot)$ does not change sign while on the other hand the antisymmetry requires
$g(\rho_1,\rho_2) \kappa_2(\rho_2) = - g(\rho_2,\rho_1) \kappa_1(\rho_2)$
which is contradictory since $\kappa_1$ and $\kappa_2$ are both strictly positive.
How the current symmetry rules out specific microscopic dynamics for a product measure 
is demonstrated for a concrete example in the Appendix \ref{app:CurrentSymmetry}.

\section{Dynamical universality classes for two conservation laws}

In the remainder of this paper we restrict ourselves to two conservation laws and discuss criteria for the occurrence of the golden and other universality classes. To this end, the mode coupling 
coefficients $G^\alpha_{\beta\beta}$
are computed in terms of the matrix elements of $K$ and $J$.
Related results of \cite{Popk15b,Spoh15} do not allow for immediate predictions 
from the stationary currents for a given particle
system since in these works the mode coupling matrices are not
directly expressed in term of the stationary currents. This is achieved
below.

\subsection{Eigenmodes}

The system of stochastic Burgers equations \eref{CSBE} is expressed
in terms of eigenmodes for which the current Jacobian $J$ is diagonal.
For the diagonalization of $J$ we use as parameters
the reduced trace $\theta$, the asymmetry $\omega$, and the signed square root $\delta$ 
of the discriminant which are defined by
\bel{thetadeltadef}
\theta := J_{11} - J_{22}, \quad 
\omega := \frac{J_{12}}{J_{21}}, \quad
\delta := \left\{\ba{ll}
\theta \sqrt{1 + \frac{4 J_{12} J_{21}}{\theta^2}} 
& \theta \neq 0 \\[2mm]
\sqrt{4 J_{12} J_{21}} 
& \theta = 0 .
\ea \right.
\ee
Only current Jacobians
with non-degenerate eigenvalues
\bel{eval} 
\lambda_{\pm} 
= \frac{1}{2} \left( J_{11} + J_{22} \pm \delta \right),
\ee 
i.e., with non-zero discriminant, are considered below which ensures a strictly hyperbolic conservation law \eref{hscl}.

%The transposition of a matrix is denoted by the superscript $T$.
The following proposition concerning the
diagonalization of a $2\times 2$ matrix is high school math. It serves to introduce
notation for the eigenmodes and some of their basic properties.

\begin{prop}
Let $j^\alpha$ be the stationary currents of a  two-component system 
satisfying the current symmetry \eref{currsym} with 
compressibility matrix $K$ \eref{Kdef} with strictly positive diagonal elements
and current Jacobian $J$  with nonvanishing discriminant.
With the functions
\bel{uvdef}
u_+ := \frac{\delta - \theta}{2 J_{21}}, \quad 
u_- := - \frac{u_+}{\omega} , \quad 
v := 1-u_+u_-  ,
\ee
the normalization factors
\bel{ypmdef}
z_+^2  := \kappa_1 + 2 u_+ \bar{\kappa} + u_+^2 \kappa_2, \quad 
z_-^2 :=  u_-^2 \kappa_1 + 2 u_- \bar{\kappa} + \kappa_2,
\ee
and the matrix elements
\bea 
& & u_1^+ := \frac{1}{z_+}, \quad
u_2^+ := \frac{1}{z_+} u_+, \quad
u_1^- := \frac{1}{z_-} u_-, \quad
u_2^- := \frac{1}{z_-}
\label{levec12} \\[2mm]
& & v_1^+ := \frac{z_+}{v} , \quad
v_2^+ :=  - \frac{z_+}{v} u_-, \quad
v_1^- := - \frac{z_-}{v} u_+, \quad
v_2^- :=  \frac{z_-}{v}   
\label{revec12} 
\eea
the matrices
\bel{Rdef}
R := \left( \ba{cc} u_1^+ & u_2^+ \\ u_1^- & u_2^- \ea \right), \quad
R^{-1} := \left( \ba{cc} v_1^+ & v_1^- \\ v_2^+ & v_2^- \ea \right)
\ee
satisfy 
\bel{ortho}
\mathds{1} = R R^{-1} = R K R^T, \quad
R J R^{-1} = \left( \ba{cc} \lambda_+ & 0 \\ 0 & \lambda_- \ea \right) 
\ee 
where $\mathds{1}$ is the two-dimensional unit matrix.
\end{prop}

\proof The matrices $R, R^{-1}$ read in terms of the quantities $z_\pm $, $u_\pm$, and $v$
\be
R = \left( \ba{cc} z_+^{-1} & z_+^{-1} u_+ \\ z_-^{-1} u_- & z_-^{-1} \ea \right), \quad
R^{-1} = v^{-1}\left( \ba{cc} z_+ & - z_-  u_+ \\ - z_+  u_- &  z_- \ea \right).\nonumber
\ee

The definition \eref{uvdef} of $v$
yields immediately the first equality in \eref{ortho}.
By the definitions \eref{ypmdef} of the normalization factors the diagonal elements of $R K R^T$ are equal to 1.
The function $u_+$ satisfies the quadratic equation
\be
J_{21} u_+ - J_{12} u_+^{-1} +  \theta = 0\nonumber
% u_+^2 =  (J_{12} -  \theta u_+)/ J_{21}
\ee
which implies the two equalities
\be 
1 + u_+ u_- = \frac{\theta u_+}{J_{12}}, \quad
1 - u_+ u_- = \frac{J_{12}}{\delta u_+}. \nonumber
\ee
The first equality together with the current symmetry 
\eref{currsym2} yields
$(R K R^T)_{12} = (R K R^T)_{21} =0$.
The second equality proves the diagonalization of $J$.
\qed

\begin{cor}
The column and row vectors 
\be
\vec{v}_\pm := \left( \ba{c} v_1^\pm \\ v_2^\pm \ea \right), \quad 
\vec{u}_\pm := \left( u_1^\pm , \, u_2^\pm \right).\nonumber
\ee
are right and left eigenvectors of $J$ with
eigenvalues $\lambda_\pm$ and they
satisfy the biorthogonality relation
\be
\vec{u}_s \cdot \vec{v}_{s'} := u_1^s v_1^{s'} + u_2^s v_2^{s'} =
\delta_{s,s'}.\nonumber
\ee
for $s,s'\in\{+,-\}$.
\end{cor}

%\begin{rem}
Since $\delta > 0$ by assumption, the current Jacobian is diagonalizable for all 
permissible parameter values.
In particular, the tridiagonal case $J_{12} J_{21} =0$ is covered by the proposition
by taking appropriate limits: For $J_{12} =0$ one gets $\delta = \theta$, $u_+ = 0$, $u_- = - J_{21}/\theta$, while for $J_{21} =0$
one has $\delta = - \theta$, $u_+ = - J_{12}/\theta$, $u_- = 0$.

\subsection{Mode-coupling matrices}

To detail the conditions that lead to the various dynamical universality classes
predicted by mode coupling theory we express in explicit form
the diagonal elements of the mode coupling
matrices in terms of the stationary currents and compressibilities.

\begin{df}
\label{Def:G}
The mode-coupling matrices $G^\gamma$ are the linear combinations
\bel{Gdef}
G^\gamma := \frac{1}{2} \sum_{\lambda} R_{\gamma \lambda}  
(R^{-1})^T H^{\lambda} R^{-1} 
\ee
of transformed Hessians $H^{\lambda}$.
\end{df}
They satisfy $G^\gamma_{\alpha\beta} = G^\gamma_{\beta\alpha}$ since the
Hessians $H^{\lambda}$ are symmetric by construction.  It is also possible  to express the Hessians
in terms of  the mode-coupling matrices,  see Appendix~\ref{app:HviaG}.

According to mode coupling theory only the diagonal elements
of the mode coupling matrices determine the large-scale
behaviour of the fluctuations. Hence we ignore the off-diagonal
elements and we also recall that the normalization factors
$z_\pm$ of the eigenvectors are non-zero for any permissible choice of
parameters. Therefore the ratios
\be 
g:= \frac{\sqrt{z_+z_-}}{v^2}, \quad y := \sqrt{\frac{z_+}{z_-}} 
\ee
are non-zero and well-defined. The theorem below  expresses
the diagonal elements $G^\alpha_{\beta\beta}$ of the mode coupling
matrices in terms of the stationary currents and compressibilities.
We stress that the compressibilities enter only the non-zero overall amplitude
given in terms of $g$ and $y$. Whether a matrix element vanishes
or not is solely determined by the current-density relation via the
Hessians and the unnormalized eigenvector components of the current
Jacobian.

\begin{theo}
The diagonal elements of the mode coupling matrices are given by
\bea 
G^1_{11} 
& = & \frac{g}{2} y \left[ H^1_{11} + u_+ H^2_{11} 
- 2 u_- (H^1_{12} + u_+ H^2_{12})
+ u_-^2 (H^1_{22} + u_+ H^2_{22}) \right]
\label{G111}\\
G^1_{22} 
& = & \frac{g}{2}  y^{-3} \left[u_+^2 (H^1_{11} + u_+ H^2_{11}) 
- 2 u_+ (H^1_{12} + u_+ H^2_{12}) + H^1_{22} 
+ u_+  H^2_{22} \right] %\nonumber \\
\label{G122} \\
G^2_{11} 
& = & \frac{g}{2}  y^3 \left[u_- H^1_{11} + H^2_{11}
- 2 u_- (u_- H^1_{12} + H^2_{12}) 
+ u_-^2 (u_- H^1_{22} + H^2_{22})\right] 
\label{G211} \\
G^2_{22} 
& = & \frac{g}{2}  y^{-1} \left[u_+^2 (u_- H^1_{11} + H^2_{11})
- 2 u_+ (u_- H^1_{12} + H^2_{12})
+ u_- H^1_{22} + H^2_{22}\right] 
\label{G222}
\eea
\end{theo}

\proof To compute $G^\gamma$ we note that with the diagonal normalization matrix
\be
Z = \left( \ba{cc} z_+ & 0 \\ 0 & z_- \ea \right)\nonumber
\ee
and the rescaled diagonalization matrices
\be
\tilde{R} = \left( \ba{cc} 1 & u_+ \\ u_- & 1 \ea \right), \quad
\tilde{R}^{-1} = v^{-1}\left( \ba{cc} 1 & - u_+ \\ - u_- & 1 \ea \right)\nonumber
\ee
one has $R = Z^{-1} \tilde{R}$. Defining furthermore
\be 
S := v \tilde{R}^{-1} = \left( \ba{cc} 1 & - u_+ \\ - u_- & 1 \ea \right)\nonumber
\ee
yields the modified mode coupling matrix
\be
\tilde{G}^\gamma :=
 2 v^2  Z^{-1} G^\gamma Z^{-1} = \sum_{\lambda} R_{\gamma \lambda}  
S^T H^{\lambda} S \nonumber
\ee
with matrix elements
\bea 
& & \tilde{G}^\gamma_{11} = \frac{2 v^2}{z_+^2} G^\gamma_{11}, \quad 
\tilde{G}^\gamma_{22} = \frac{2 v^2}{z_-^2} G^\gamma_{22} \nonumber\\
& & \tilde{G}^\gamma_{12} = \frac{2 v^2}{z_+ z_-} G^\gamma_{12}, \quad 
\tilde{G}^\gamma_{21} = \frac{2 v^2}{z_+ z_-} G^\gamma_{21}.\nonumber
\eea

% In particular, we note that $\tilde{G}^\gamma_{\alpha\beta} = 0$ if and only iff $G^\gamma_{\alpha\beta} = 0$.

The matrix multiplication can now be performed in a straightforward fashion and
yields the matrix elements
\bea 
(S^T H^{\lambda} S)_{11} & = & H^\lambda_{11} - 2 u_- H^\lambda_{12} + u_-^2 H^\lambda_{22} \nonumber\\
(S^T H^{\lambda} S)_{12} & = & (1+u_+ u_-) H^\lambda_{12}- u_+ H^\lambda_{11} - u_- H^\lambda_{22} \nonumber\\
(S^T H^{\lambda} S)_{22} & = & H^\lambda_{22} - 2 u_+ H^\lambda_{12} + u_+^2 H^\lambda_{11}\nonumber
\eea
and $(S^T H^{\lambda} S)_{21}=(S^T H^{\lambda} S)_{12}$ by symmetry.
Using the definition \eref{Gdef} therefore gives
\bea 
G^1_{11} 
& = & \frac{z_+}{2 v^2} \left[H^1_{11} - 2 u_- H^1_{12} + u_-^2 H^1_{22}
+ u_+ \left(H^2_{11} - 2 u_- H^2_{12} + u_-^2 H^2_{22}\right)\right] 
\nonumber \\
G^1_{22} 
& = & \frac{z_-^2}{2 v^2 z_+} \left[H^1_{22} - 2 u_+ H^1_{12} + u_+^2 H^1_{11} 
+ u_+ \left(H^2_{22} - 2 u_+ H^2_{12} + u_+^2 H^2_{11}\right)\right] 
\nonumber\\
G^2_{11} 
& = & \frac{z_+^2}{2 v^2 z_-} \left[u_-
\left(H^1_{11} - 2 u_- H^1_{12} + u_-^2 H^1_{22}\right)
+ H^2_{11} - 2 u_- H^2_{12} + u_-^2 H^2_{22}\right] 
\nonumber\\
G^2_{22} 
& = & \frac{z_-}{2 v^2} \left[u_- \left(H^1_{22} - 2 u_+ H^1_{12} + u_+^2 H^1_{11}\right)
+ H^2_{22} - 2 u_+ H^2_{12} + u_+^2 H^2_{11}\right] .\nonumber
\eea
Regrouping terms yield \eref{G111} - \eref{G222}.
\qed

\subsection{Scenarios for dynamical universality classes}

Generically, i.e., for two vanishing self-coupling coefficients $G^{1}_{11} \neq 0$ and $G^{2}_{22} \neq 0$
mode coupling theory predicts two KPZ modes. To search for the golden universality class we consider 
the mode coupling scenarios
with less than two KPZ modes in more detail. We note that $u_+ u_- \neq 1$
since $v = 1 - u_+ u_-  \neq 0 $ by construction. It follows that
$u_-^{-1}\neq u_+$. 

\subsubsection{A) One vanishing self-coupling coefficient}

The following cases are permitted by mode coupling theory.\\

\noindent \underline{(i) Mode 1 is KPZ and mode 2 is 5/3-L\'evy:}
This requires $G^1_{11} \neq 0$, $G^2_{22}=0$, $G^2_{11} \neq 0$, with arbitrary $G^1_{22}$.
Thus 
\bea 
2 (H^1_{12} + u_+ H^2_{12}) & \neq & u_-^{-1} (H^1_{11} + u_+ H^2_{11})
+ u_-  (H^1_{22} + u_+ H^2_{22}), 
\nonumber\\
0 
& \neq & \omega (u_- H^1_{11} + H^2_{11})
+ (u_- H^1_{22} + H^2_{22}),
\nonumber \\
2 (u_- H^1_{12} + H^2_{12})
& = & u_+ (u_- H^1_{11} + H^2_{11})
+ u_+^{-1} (u_- H^1_{22} + H^2_{22}).
\nonumber
\eea

\noindent \underline{(ii) Mode 1 is KPZ and mode 2 is diffusive:}
This requires $G^1_{11} \neq 0$, $G^1_{22}=G^2_{22}=G^2_{11}=0$. Thus
\bea 
H^1_{22} + u_+  H^2_{22}
& \neq &  - \omega (H^1_{11} + u_+ H^2_{11}),
\nonumber\\
2 (H^1_{12} + u_+ H^2_{12})
& = & u_+ (H^1_{11} + u_+ H^2_{11}) 
+ u_+^{-1} (H^1_{22} + u_+  H^2_{22}),
\nonumber \\
u_- H^1_{22} + H^2_{22}
& = & - \omega (u_- H^1_{11} + H^2_{11}),
\nonumber \\
2  (u_- H^1_{12} + H^2_{12})
& = & (u_+ + u_-^{-1}) (u_- H^1_{11} + H^2_{11}).
\nonumber
\eea

\noindent \underline{(iii) Mode 1 is modified KPZ and mode 2 is diffusive:}
This requires $G^1_{11} \neq 0$, $G^1_{22}\neq 0$, and $G^2_{22}=G^2_{11}=0$. Thus
\bea 
2 (H^1_{12} + u_+ H^2_{12}) 
& \neq & u_-^{-1} (H^1_{11} + u_+ H^2_{11})
+ u_- (H^1_{22} + u_+ H^2_{22}), 
\nonumber\\
2 (H^1_{12} + u_+ H^2_{12})  
& \neq & u_+ (H^1_{11} + u_+ H^2_{11}) 
+ u_+^{-1} (H^1_{22} + u_+  H^2_{22}),
\nonumber \\
u_- H^1_{22} + H^2_{22}
& = & - \omega (u_- H^1_{11} + H^2_{11}),
\nonumber \\
2  (u_- H^1_{12} + H^2_{12})
& = & (u_+ + u_-^{-1}) (u_- H^1_{11} + H^2_{11}).
\nonumber
\eea

\subsubsection{B) Two vanishing self-coupling coefficients}

This requires $G^2_{22}=G^1_{11}=0$ and therefore
\bea 
2 (H^1_{12} + u_+ H^2_{12})
& = &  u_-^{-1} (H^1_{11} + u_+ H^2_{11})
+ u_- (H^1_{22} + u_+ H^2_{22}) 
\label{G111noKPZ}\\
2 (u_- H^1_{12} + H^2_{12})
& = & u_+ (u_- H^1_{11} + H^2_{11})
+ u_+^{-1} (u_- H^1_{22} + H^2_{22}) .
\label{G222noKPZ}
\eea

The remaining diagonal mode coupling coefficients take the form
\bea 
G^1_{22} 
& = & \frac{z_-^2}{2 v z_+} \left[ \omega (H^1_{11} + u_+ H^2_{11}) 
+ (H^1_{22} + u_+  H^2_{22}) \right]  
\label{G122noKPZ} \\
G^2_{11} 
& = & \frac{z_+^2}{2 v z_- \omega} \left[\omega (u_- H^1_{11} + H^2_{11})
+ (u_- H^1_{22} + H^2_{22})\right] .
\label{G211noKPZ} 
\eea

\noindent \underline{(i) Two modes are golden L\'evy:}
\bea 
0 & \neq & \omega (H^1_{11} + u_+ H^2_{11}) + (H^1_{22} + u_+  H^2_{22})
\label{G122GM} \\
0  & \neq & \omega (u_- H^1_{11} + H^2_{11}) + (u_- H^1_{22} + H^2_{22}) .
\label{G211GM} 
\eea

\noindent \underline{(ii) Mode 1 is 3/2-L\'evy and mode 2 is EW:}
\bea 
0 & \neq & \omega (H^1_{11} + u_+ H^2_{11}) + (H^1_{22} + u_+  H^2_{22})
\label{G12232LD} \\
0  & = & \omega (u_- H^1_{11} + H^2_{11}) + (u_- H^1_{22} + H^2_{22}) .
\label{G21132LD} 
\eea

\noindent \underline{(iii) Two modes are EW:}
\bea 
0 & = & \omega (H^1_{11} + u_+ H^2_{11}) + (H^1_{22} + u_+  H^2_{22})
\label{G122DD} \\
0  & = & \omega (u_- H^1_{11} + H^2_{11}) + (u_- H^1_{22} + H^2_{22}) .
\label{G211DD} 
\eea

We conclude that the conditions \eref{G111noKPZ}, \eref{G222noKPZ},
\eref{G122GM}, and \eref{G211GM} must be met for the occurrence of the
golden ratio. In this case both modes are golden L\'evy.

\begin{rem}
The current symmetry imposes the constraints
\bea
H^1_{12} \det{(K)} 
& = & H^2_{11} \kappa_1^2 - H^1_{11} \kappa_1 \bar{\kappa}
+ H^1_{22} \bar{\kappa} \kappa_2 - H^2_{22} \bar{\kappa}^2
\nonumber \\ & & 
 + J_{21} (\kappa_1 \partial_{1}
- \bar{\kappa} \partial_{2}) \kappa_1
 - J_{12} (\kappa_1 \partial_{1} 
- \bar{\kappa} \partial_{2}) \kappa_2 
\nonumber \\ & & 
- (J_{11} - J_{22}) (\kappa_1 \partial_{1}
- \bar{\kappa} \partial_{2} ) \bar{\kappa}
\label{currsym1}
\eea
and
\bea
H^2_{12} \det{(K)}
& = & H^2_{11} \bar{\kappa} \kappa_1 - H^1_{11} \bar{\kappa}^2 + H^1_{22} \kappa_2^2 - H^2_{22} \bar{\kappa} \kappa_2 
\nonumber \\ & & 
+  J_{12} (\kappa_2 \partial_{2}
 - \bar{\kappa} \partial_{1}) \kappa_2
- J_{21} (\kappa_2 \partial_{2} - \bar{\kappa} \partial_{1}) \kappa_1
\nonumber \\ & & 
+ (J_{11} - J_{22}) ( \kappa_2 \partial_{2}
- \bar{\kappa} \partial_{1}) \bar{\kappa}
\label{currsym2}
\eea
on the matrix elements of the Hessians. Hence the diagonal elements \eref{G111} - \eref{G222} of the mode coupling matrix and therefore the mode scenarios discussed above can be expressed entirely in terms of $J$, $K$, and the diagonal elements of the Hessians rather than
in terms of $J$, $K$, and the full Hessians.
\end{rem}

\section{Symmetric or antisymmetric currents at equal densities}

%\section{Applications}

The golden universality class has been observed in  \cite{Popk15b,Spoh15}
on a special parameter manifold that requires fine-tuning of densities and/or interaction parameters. We investigate here whether it can occur in more generic circumstances, viz., for the symmetry property of the current-density relation
$j^2(\rho_1,\rho_2) = \pm j^1(\rho_2,\rho_1)$ at the point of equal densities. 
$\rho_1=\rho_2$ where
\bea 
 J_{11} & = & \pm  J_{22} \\
 J_{12} & = & \pm J_{21} .
\eea
These properties of $J$ are sufficient to show which mode coupling coefficients vanish.

\subsection{Antisymmetric case}

In the antisymmetric case one has for equal densities $\theta\neq 0$ and $\omega =-1$
which gives $u_- = u_+$ and $\lambda_- = - \lambda_+$. 
The current symmetry implies $\bar{\kappa} \neq 0$ which means that a microscopic physical
dynamics
with a partially factorized invariant measure of the form \eref{pfim}
and antisymmetric current-density-relation at equal densities does not exist.

However, non-vanishing cross correlations between the conserved densities are not ruled out.
The Hessians then have the symmetry relations
\be
H^2_{11}  =  - H^1_{22}, \quad
H^2_{12}  =  - H^1_{12},\quad
H^2_{22}  =  - H^1_{11}.\nonumber
\ee
This yields the mode coupling coefficients
\bea 
G^1_{11} 
& = & \frac{u_+ z_+}{2 v^2} \left[u_+^{-1}(1 - u_+^3)  H^1_{11}
- 2 (1 - u_+) H^1_{12}  - (1 - u_+) H^1_{22}
\right]
\nonumber\\
G^1_{22} 
& = & \frac{u_+ z_-^2 }{2 v^2 z_+} \left[- (1 - u_+) H^1_{11}
- 2 (1 - u_+) H^1_{12} + u_+^{-1}(1 - u_+^3)H^1_{22} \right] 
\nonumber\\
G^2_{11} 
& = & - \left(\frac{z_+}{z_-}\right)^3 G^1_{22} 
\nonumber \\
G^2_{22} 
& = & - \frac{z_-}{z_+} G^1_{11} 
\nonumber
\eea
Hence the following scenarios are possible: \\

\noindent \underline{(i) 2 KPZ modes:}
$u_+\neq 1$ and
\be 
u_+ (H^1_{22} + 2 H^1_{12}) \neq (1 + u_+ + u_+^2) H^1_{11} 
\ee
\underline{(ii) 2 golden modes:} 
$u_+\neq 1$ and
\bea 
u_+ (H^1_{22} + 2 H^1_{12}) & = & (1 + u_+ + u_+^2) H^1_{11} \\
H^1_{11} & \neq & H^1_{22} 
\eea
\noindent \underline{(iii) 2 EW modes:} All diagonal mode coupling coefficients vanish
if either $u_+=1$ or
\bea 
2 u_+ H^1_{12} & = & (1 +  u_+^2) H^1_{11} \\
H^1_{11} & = & H^1_{22} 
\eea
In all allowed cases both modes are the same.
These four cases correspond to the diagonal of the mode coupling table
of \cite{Popk15b}.

\subsection{Symmetric case}

In the symmetric case the properties of the Jacobian correspond to $\theta = 0$ and $\omega=1$
and therefore $u_\pm = \pm 1$. The non-degeneracy of the eigenvalues $\lambda_\pm$ implies
$J_{12} J_{21} \neq 0$.
There is no conflict with the current symmetry.
The Hessians have the symmetry relations
\be
H^2_{11}  =  H^1_{22}, \quad
H^2_{12}  =  H^1_{12}, \quad
H^2_{22}  =  H^1_{11}\nonumber
\ee
From \eref{G111} - \eref{G222} one gets
\bea 
G^1_{11} 
& = & \frac{1}{4 (z_+^{-1}) } \left[H^1_{11} + H^1_{22}
+ 2 H^1_{12} \right], 
\nonumber\\
G^1_{22} 
& = & \frac{(z_+^{-1})}{4 (z_-^{-1})^2} \left[H^1_{11} +  H^1_{22}
- 2 H^1_{12}\right],
\nonumber \\
G^2_{11} &=& G^2_{22}
 =  0. \nonumber
\eea
Hence in all three cases mode 2 is diffusive, thus ruling out the golden mode. Mode 1 can be KPZ, mKPZ, 3/2L, or EW.
These four cases correspond to the bottom row (rightmost column) of the mode coupling table of \cite{Popk15b}.

\section{Conclusions}

So far the golden universality class for fluctuations
in driven systems with two conservation laws has remained somewhat elusive.
It was only found by fine-tuning of the conserved densities or interaction
parameters which determine the form of the stationary current-density relations
\cite{Popk15b,Spoh15}.
In the quest for a natural occurrence that does not require fine-tuning we have shown that systems with currents that are
symmetric at equal densities cannot exhibit golden modes. In fact, in this case at least
one of the two mode is diffusive with dynamical exponent $z=2$ and belongs to the EW universality whereas the other mode
is KPZ, modified KPZ or 3/2-L\'evy  with dynamical exponent $z=3/2$ in each case.

However, the golden modes can appear with systems with two conservation laws
if the currents are antisymmetric at equal densities. In this case, both modes belong
to the same universality class which may be KPZ, EW or golden. However, such systems
cannot have invariant measures that factorize over the two conserved quantities.
This property may explain the rarity of analytical and simulation results for the golden modes, as 
one often considers such systems to make them amenable to explicit computations.
On the other hand, from a physical perspective such a factorization is highly exceptional which gives hope that the golden universality class might be less unusual as it appears at the present state of the art.

The nonexistence of physical systems with antisymmetric currents at equal densities
and an invariant measure that factorizes over the conserved quantities
derives from
the current symmetry that arises like Onsager's reciprocity relations
from time-reversal.
Thus the current symmetry is not only of fundamental significance
but also helps ruling out microscopic dynamics in the search for other
modes without having to check whether an invariant measure that is amenable to exact computation of the current is invariant for a candidate microscopic dynamics. Potential applications include
two-lane versions of zero-range processes \cite{Gros03} or KLS-models \cite{Katz84,KLS,Hage01,Dier13}
which are exclusion processes with next-nearest-neighbour interactions
with the invariant measure of a one-dimensional Ising model for each lane.

\begin{acknowledgments}
It is a pleasure to acknowledge stimulating discussions with P. Gon{\c c}alves. This work is financially supported by FCT/Portugal through CAMGSD, IST-ID, Projects UIDB/04459/2020 and UIDB/04459/2020 and by the FCT Grant 2022.09232.PTDC.
 V.~P. acknowledges  financial support from the Deutsche Forschungsgemeinschaft through DFG
 project KL 645/20-2,  and thanks IST Lisbon for hospitality and support. 
\end{acknowledgments}

\appendix

\section{Current symmetry and constraints on the microscopic dynamics}
\label{app:CurrentSymmetry}

We illustrate how the current symmetry rules out microscopic dynamics for some two-lane exclusion processes. These driven lattice gases are defined on
two one-dimensional lattices (that we call lanes), that are indexed by $\alpha \in \{1,2\}$
and have $L$ sites each. We view the two lanes as being arranged in parallel, with a pair of sites on the two lanes labeled by an integer $k$.

%\subsection{Two-lane exclusion process}
%
%For the two-lane exclusion process \cite{Popk04} discussed above the
%currents \eref{twolanecurrents} with $b:=b_1/2=b_2/2$ as required
%by the current symmetry and with different
%asymmetries in each lane one gets for the stationary currents
%\bel{twolanecurrents}
%j^1(\rho_1,\rho_2) = j(\rho_1)(f_1 + b \rho_2), \quad
%j^2(\rho_1,\rho_2) = j(\rho_2)(f_2 + b \rho_1)
%\ee
%where $j(x) = x(1-x)$. The static compressibilities are given by $\kappa_\alpha(\rho_\alpha) =
%j(\rho_\alpha)$. With $j'(x) = 1 - 2x$ the current Jacobian is 
%therefore given
%by
%\be
%J = \left( \ba{cc} j'(\rho_1)(f_1+b \rho_2) & b j(\rho_1) \\ 
%b j(\rho_2) & j'(\rho_2)(f_2+b \rho_1) \ea \right) 
%\ee
%which yields
%\bea 
%\theta & = & f_1 (1-2\rho_1) - f_2 (1-2\rho_2) + b (\rho_2 -\rho_1) \\
%\eea
%The Hessians read
%\be
%H^1 = \left( \ba{cc} -2 (f_1+b \rho_2) & b j'(\rho_1) \\ 
%b j'(\rho_1) & 0 \ea \right), \quad
%H^2 = \left( \ba{cc} 0 & b j'(\rho_2) \\ 
%b j'(\rho_2) & -2 (f_2+b \rho_1) \ea \right)
%\ee
%
%The conditions \eref{G111noKPZ}, \eref{G222noKPZ},
%\eref{G122GM}, and \eref{G211GM} for the occurrence of the
%golden ratio which appears necessarily in both modes 
%thus read
%\bea 
%2 (H^1_{12} + u_+ H^2_{12})
%& = & - u_+^{-1} \frac{J_{12}}{J_{21}} H^1_{11} 
%- \frac{J_{21}}{J_{12}} u_+^2 H^2_{22}
%\label{G111noKPZTwolaneASEP}\\
%2 (- u_+ \frac{J_{21}}{J_{12}} H^1_{12} + H^2_{12})
%& = & - u_+^2 \frac{J_{21}}{J_{12}} H^1_{11} 
%+ u_+^{-1} H^2_{22} .
%\label{G222noKPZTwolaneASEP}
%\eea
%
%
%

%For the two-lane exclusion process \cite{Popk04}
Particles jump 
along each lane (but not to the other) obeying the exclusion rule, i.e., a jump attempt fails if the target site is already occupied.  Thus the state of each lattice site $k$ is described by a pair of occupation numbers $n^\alpha_k \in \{0,1\}$, indicating whether lane $\alpha$ is
occupied by a particle or not.

To be definite, we look for jump dynamics such that the NESS has a particularly simple form given by an
invariant measure that is a product measure factorizing over all sites of the two lattices, i.e., 
\bel{invmeastwolaneASEP} 
\pi_{\rho_1,\rho_2}(n^1_1,\dots,n^1_L,n^2_1,\dots,n^2_L) = \prod_{\alpha=1}^2 \prod_{k=1}^L [(1-\rho_\alpha)(1-n^\alpha_k) + \rho_\alpha n^\alpha_k]
\ee 
with average density $\rho_\alpha$ on each lane, compressibilities
$\kappa_\alpha=\rho_\alpha(1-\rho_\alpha)$ and no correlations
between different sites which implies $\bar{\kappa}=0$.
Well-studied examples for single-lane dynamics with these properties 
dynamics include the paradigmatic asymmetric simple exclusion process \cite{Derr98,Ligg99,Schu01}, or the $k$-step exclusion processes with long-range jumps
\cite{Guio99,Guio02}, to mention just a few. 

We denote the stationary currents arising
from these dynamics in the absence of any coupling between the lanes as
$j^\alpha_0(\rho_\alpha)$. It is well-known that fluctuations in such systems
generically belong to the KPZ universality class \cite{Halp15}. To study non-trivial coupled systems that may exhibit other fluctuation patterns we allow the
transition rates in each lane to depend on the occupation numbers of the other lane
and ask whether such models with the invariant product measure \eref{invmeastwolaneASEP} can exist. 

To be specific, we consider three cases:\\
(i) Totally asymmetric dynamics with equal dynamics on both lanes
such that the rate 
for a forward jump from site $k$ to a site $k+r$ on lane $1$ is enhanced by the
number of particles on sites $k$ and $k+r$ on lane 2 and the rate of a 
backward jump from site $k$ to a site $k-r$ on lane $2$ is enhanced by the
number of particles on sites $k$ and $k-r$.\\
(ii) Totally asymmetric dynamics with equal dynamics on both lanes
but opposite
directionality
such that the rate 
for a forward jump from site $k$ to a site $k+r$ on lane $1$ is enhanced by the
number of particles on sites $k$ and $k+r$ on lane 2 and the rate of a 
backward jump from site $k$ to a site $k-r$ on lane $2$ is enhanced by the
number of particles on sites $k$ and $k-r$.\\
(iii) Facilitated totally asymmetric dynamics, with parallel directionality 
on both lanes and rates such that 
a forward jump from site $k$ to a site $k+r$ on lane $\alpha$ is enhanced by a factor $b_\alpha$ when a site $k'$ is occupied on both lanes (and 
$k'$ different from $k$ and the target site 
$k+r$).

In case (i) the factorization of the invariant measure yields the 
exact currents
\bel{twolanecurrents}
j^1(\rho_1,\rho_2) = j_0(\rho_1)(1 + 2 b_1 \rho_2), \quad
j^2(\rho_1,\rho_2) = j_0(\rho_2)(1 + 2 b_2 \rho_1)
\ee
and therefore $J_{12}(\rho_1,\rho_2) = 2 b_1 j^1_0(\rho_1)$ and $J_{21}(\rho_1,\rho_2) = 2 b_2 j^2_0(\rho_2)$. For such dynamics the current symmetry 
\eref{currsymfac} requires
\bel{currsymfac1}
b_2 j_0(\rho_2) \rho_1(1-\rho_1) = b_1 j_0(\rho_1) \rho_2(1-\rho_2).
\ee
Therefore such dynamics exist if and only if $b_1 = b_2$ and
$j_0(x) = c x(1-x)$ as in the conventional TASEP. For $r=1$ this is 
the two-lane model of \cite{Popk04}. Once still needs to prove
that for $b_1 = b_2$ the product measure \eref{invmeastwolaneASEP} is invariant
(as is done in \cite{Popk04}) but there is no reason to check invariance in the
more general case $b_1 \neq b_2$: The macroscopic current symmetry
asserts that such a microscopic model does not exist.

In case (ii) the factorization of the invariant measure yields the 
exact currents
\be 
j^1(\rho_1,\rho_2) = j_0(\rho_1)(1 + 2 b_1 \rho_2), \quad
j^2(\rho_1,\rho_2) = - j_0(\rho_2)(1 + 2 b_2 \rho_1)
\ee
and therefore $J_{12}(\rho_1,\rho_2) = 2 b_1 j^1_0(\rho_1)$ and $J_{21}(\rho_1,\rho_2) = - 2 b_2 j^2_0(\rho_2)$. For such dynamics to exist the current symmetry 
\eref{currsymfac} requires
\be
b_2 j_0(\rho_2) \rho_1(1-\rho_1) = - b_1 j_0(\rho_1) \rho_2(1-\rho_2).
\ee
Therefore such dynamics exist if and only if $b_1 = - b_2$ and
$j_0(x) = c x(1-x)$ as in the conventional TASEP. Hence a model with current
enhancement on both lanes due the presence of particles on the other lane
need not be considered.
The macroscopic current symmetry
asserts that such a microscopic model does not exist.

In case (iii) the factorization of the invariant measure yields the 
exact currents
\be 
j^1(\rho_1,\rho_2) = j^1_0(\rho_1)(1 + b_1 \rho_1 \rho_2), \quad
j^2(\rho_1,\rho_2) = j^2_0(\rho_2)(1 + b_2 \rho_1 \rho_2)
\ee
Thus $J_{12}(\rho_1,\rho_2) = b_1 \rho_1 j^1_0(\rho_1)$ and $J_{21}(\rho_1,\rho_2) = b_2 \rho_2 j^2_0(\rho_2)$. For such dynamics to exist the current symmetry 
\eref{currsymfac} requires
\be 
b_2 j^2_0(\rho_2) (1-\rho_1) = b_1 j^1_0(\rho_1) (1-\rho_2).
\ee
This is valid for all $(\rho_1,\rho_2)$ if and only if $b_1=b_2=0$, 
i.e., no such coupled facilitated dynamics for which the fully factorized measure is invariant exists.

\section{On the reliability of mode coupling theory for nonlinear fluctuating hydrodynamics}
\label{app:NLIEforOscillators}

Consider an infinite chain of particles of unit mass at positions $q_{j}$
interacting through a pair potential $V(r_{j})$ where $r_{j}=q_{j} - q_{j-1}$ 
is interparticle distance. With the momentum $p_{j}$ of the particles
the total energy is given by
\be 
E = \sum_{j\in\Z} \left[\half p_{j}^2 + V(r_{j})\right]
\ee
and the deterministic Hamiltonian dynamics for such a system are given by
\bea 
\ddt r_{j}(t) & = & p_{j}(t) - p_{j-1}(t) 
\label{Hamr} \\
\ddt p_{j}(t) & = & V'(r_{j+1}(t)) - V'(r_{j}(t)).
\label{Hamp}
\eea
Evidently, besides the energy $E$ also the total momentum 
$P$ and ``volume'' $V$, i.e., the quantities
\be 
P = \sum_{j\in\Z} p_{j}, \quad V = \sum_{j\in\Z} r_{j}
\ee
are conserves under the dynamics. 

For the existence of such dynamics for general potentials see 
\cite{Bern09}. In the special case of harmonic oscillators
with $V(x) = x^2/2$ the system is completely integrable. 
This is most easily seen by introducing the variables 
\be 
\eta_{2j-1}(t) := r_{j}(t), \quad \eta_{2j}(t) := p_{j}(t).
\ee
Then the two coupled equations \eref{Hamr} and 
\eref{Hamp} can be expressed as the single linear equation
\be 
\ddt \eta_{j}(t) = \eta_{j+1}(t) - \eta_{j-1}(t)
\label{Hamharm}
\ee
which is explicitly solvable by Fourier transformation.

To introduce noise an exchange of the variables
$\eta_{j}(t),\eta_{j+1}(t)$ at exponential random times
was introduced in \cite{Bern12} for more general potentials and studied in detail for the harmonic case in \cite{Bern16}.
This random exchange violates momentum conservation
but leaves the total energy $E$ and the pseudo volume $W$, expressed in terms of the variables $\eta_j$ as
\be
E = \sum_{j\in\Z} \eta^2_j , \quad 
W = \sum_{j\in\Z} \eta_j,
\ee
are conserved.

To write the generator of the stochastic dynamics
for this randomized chain of harmonic oscillators we denote
by $\bfeta := (\dots, \eta_{j-1}, \eta_{j}, \eta_{j+1}, \dots)$
the infinite set of variables $\eta_{j}\in\R$ and 
introduce the swapped state variable $\bfeta^{kk+1}$
by the local states variables
\be 
\eta^{kk+1}_l = \eta_{l} + (\eta_{l+1} - \eta_{l})
\delta_{l,k} +  (\eta_{l-1} - \eta_{l})
\delta_{l,k+1}.
\ee
which the state reached from $\bfeta$ after an exchange
of $\eta_k$ and $\eta_{k+1}$. 
The deterministic part of the generator that yields the
Hamiltonian dynamics \eref{Hamharm} is denoted by
$\mathcal{A}$ and the generator for the random exchange
is denoted by $\mathcal{S}$. The full generator 
$\mathcal{L}$ acting on measurable functions $f(\cdot)$ is 
then given by $\mathcal{L} = \mathcal{S} + \mathcal{A}$ with
\be 
\mathcal{S} f(\eta) = \sum_{k\in\Z} 
[f(\eta^{kk+1}) - f(\eta)], \quad
\mathcal{A} f(\eta) = \sum_{k\in\Z} 
(\eta_{k+1} - \eta_{k-1}) \frac{\partial}{\partial \eta_{k}} f(\eta).
\ee

The invariant measure of the process is a product measure
\be 
\mu(\bfeta) = \prod_{k\in\Z} I(\eta_{k})
\ee
with marginals
\be 
I(x) = \sqrt{\frac{\beta}{2\pi}} \rme^{- \frac{\beta}{2}
\left(x-\rho\right)^2}.
\ee
%\bea 
%\int_{x\in\R} \rmd x \, I(x) & = & 1 \\
%\int_{x\in\R} \rmd x \, x I(x) & = & \rho \\
%\int_{x\in\R} \rmd x \, x^2 I(x) & = & \rho^2 + \frac{1}{\beta} \\
%\int_{x\in\R} \rmd x \, x^3 I(x) & = & \rho^3 + 3 \frac{\rho}{\beta} \\
%\int_{x\in\R} \rmd x \, x^4 I(x) & = & \rho^4 +  \frac{6 \rho^2}{\beta} + \frac{3}{\beta^2} \\
%\eea
The quantities $\beta$ and $\rho$ are free parameters that
can be expressed in terms of the densities of the conserved 
quantities as
\bea
\rho_1 := \exval{\eta_k} = \rho, \quad 
\rho_2 := \exval{\eta^2_k} = \rho^2 + \frac{1}{\beta}.
\eea

The static compressibilities are given by
\bea 
& & \kappa_{11} := \exval{(\eta_k-\rho_1) W} = 
\exval{\eta_k^2} - \exval{\eta_k}^2 = \frac{1}{\beta}
= \rho_2 - \rho_1^2 \\
& & \kappa_{12} := \exval{(\eta_k-\rho_1) E} = 
\exval{\eta_k^3} - \rho_1 \rho_2 
%= \rho_1^3 + 3 \rho_1 (\rho_2 - \rho_1^2) - \rho_1 \rho_2 
= 2 \rho_1 (\rho_2  - \rho_1^2) \\
& & \kappa_{22} := \exval{(\eta^2_k-\rho_2) E} = 
\exval{\eta_k^4} - \rho_2^2 
%= \rho^4 +  \frac{6 \rho^2}{\beta} + \frac{3}{\beta^2} - \rho^4 -  \frac{2 \rho^2}{\beta}
%- \frac{1}{\beta^2} 
%= \frac{4 \rho^2}{\beta} + \frac{2}{\beta^2}
= 2  (\rho_2 - \rho_1^2) \left(\rho_1^2 + \rho_2\right).
\eea
Hence the compressibilty matrix reads
\be 
K = (\rho_2 - \rho_1^2) \left(\ba{cc}
1 & 2 \rho_1 \\
2 \rho_1 & 2 (\rho_1^2+\rho_2)
\ea\right)
\ee

From the action of the generator on the locally conserved
quantities one obtains the discrete continuity equations
%\bea
%\mathcal{S} \eta_{k} & = & \eta_{k+1} + \eta_{k-1} - 2 \eta_{k} \\
%\mathcal{A} \eta_{k} & = & \eta_{k+1} - \eta_{k-1} \\
%\mathcal{S} \eta_{k}^2 & = & \eta_{k+1}^2 + \eta_{k-1}^2 - 2 \eta_{k}^2 \\
%\mathcal{A} \eta_{k}^2 & = & 2 (\eta_{k+1} - \eta_{k-1}) \eta_{k}.
%\eea
\bea
\mathcal{L} \eta_{k} & = & 2 \eta_{k+1} - 2 \eta_{k} = j^{1}_{k} - j^{1}_{k+1} \\
\mathcal{L} \eta_{k}^2 & = & \eta^2_{k+1} + \eta^2_{k-1} - 2 \eta^2_{k} + 2 (\eta_{k+1} - \eta_{k-1}) \eta_{k} = j^{2}_{k} - j^{2}_{k+1}.
\eea
with the instantaneous currents
\bea 
j^{1}_{k} = - 2 \eta_k, \quad
j^{2}_{k} = \eta^2_{k-1} - \eta^2_{k} -
2 \eta_{k-1} \eta_{k}.
\eea
The stationary currents are given by
\be 
j^1 := \exval{j^{v}_k} = - 2 \rho_1, \quad
j^2 := \exval{j^{e}_k} = - 2 \rho_1^2 
\ee 
which yields the current Jacobian
\be 
J = - 2 \left(\ba{cc}
1 & 0 \\
2 \rho_1 & 0
\ea\right).
\ee
Notice that the action of $\mathcal{L}$ on $\eta_k$
is linear and therefore the time evolution of $\rho_k(t)
:=\exval{\eta_k(t)}$
can be integrated by Fourier transformation.

Using the results derived above these stationary currents
yield the mode velocities
\be 
v_1 = -2, \quad v_2 = 0
\ee 
and the mode coupling matrices
%\paragraph{Modes:}
%Modes 
%\be 
%\vec{u}_k \equiv 
%\left(\ba{c}
%u^{v}_k  \\
%u^{e}_k 
%\ea\right) := \left(\ba{c}
%\eta_{k} - \rho  \\
%\eta_{k}^2 - \rho^2 - \beta^{-1}
%\ea\right)
%\ee
%Define
%\be 
%R := \left(\ba{cc}
%x^{-1} & 0 \\
%- 2 \rho_1 y^{-1} & y^{-1}
%\ea\right)
%\ee
%with
%\be 
%x = \frac{1}{\sqrt{\beta}}, \quad 
%y = \frac{\sqrt{2}}{\beta}
%\ee
%and notice
%\be 
%R^T = \left(\ba{cc}
%x^{-1} & - 2 \rho_1 y^{-1} \\
%0 & y^{-1}
%\ea\right), \quad
%R^{-1} = \left(\ba{cc}
%x & 0 \\
%2 \rho_1 x & y
%\ea\right), \quad
%(R^{-1})^{T} = \left(\ba{cc}
%x & 2 \rho_1 x \\
%0 & y
%\ea\right).
%\ee
%\be 
%R J R^{-1} = - 2 \left(\ba{cc}
%1 & 0 \\
%0 & 0
%\ea\right) , \quad R K R^T = \left(\ba{cc}
%1 & 0 \\
%0 & 1
%\ea\right).
%\ee
%
%
%Normal modes 
%\be 
%\vec{\phi}_k := R \vec{u}_k
%= \left(\ba{c}
%x^{-1} u^{v}_k  \\
%y^{-1} (u^{e}_k - 2 \rho_1 u^{v}_k)
%\ea\right)
%= \left(\ba{c}
%x^{-1} (\eta_{k} - \rho)  \\
%y^{-1} (\eta_{k}^2 - \rho^2 - \beta^{-1} - 2 \rho (\eta_{k} - \rho))
%\ea\right)
%\ee
%
%
%\paragraph{Mode coupling matrices:}
%Hessians
%\be
%H^{(1)} = 0, \quad 
%H^{(2)} = - 4  \left(\ba{cc}
%1 & 0 \\
%0 & 0
%\ea\right)
%\ee
%\bea 
%G^{(\alpha)} 
%& = & \half \sum_{\lambda=1}^{2} R_{\alpha\lambda} (R^{-1})^{T} H^{(\lambda)} R^{-1} \\
%& = & \half  R_{\alpha 2} (R^{-1})^{T} H^{(2)} R^{-1} 
%\eea
%
\bea 
G^{(1)} 
& = & 0 \\
G^{(2)} 
%& = & \half  y^{-1} (R^{-1})^{T} H^{(2)} R^{-1} \\
%& = & \half  x^2 y^{-1} H^{(2)}  \\
& = & - \sqrt{2} \left(\ba{cc}
1 & 0 \\
0 & 0
\ea\right).
\eea
Mode coupling theory then predicts that mode 1 is diffusive
and mode 2 belongs to the 3/2-L\'evy universality class.

To compute the scaling form of the dynamical structure
function \cite{Popk15a,Popk15b,Spoh15} one needs the
diffusion coefficient of mode 1. This can be computed directly
and rigorously from the linear evolution of the conserved
density $\eta_k(t)$ by Fourier transformation and yields
the large-scale behaviour in the moving reference frame
with velocity $v_1$ the scaling form
\be 
S_1(p,t) = \frac{1}{\sqrt{2\pi}} \rme^{- D_1 p^2 t}
\ee
with $D_1=1$. Mode coupling theory then predicts the Fourier transform of the scaling
function for mode 2 to be \cite{Popk15a}
\be 
S_2(p,t) = \frac{1}{\sqrt{2\pi}} 
\rme^{- C_0 p^{3/2} [1  - i \, \mathrm{sign}(p(v_1-v_2))] t}
\label{S2}
\ee
which is a maximally asymmetric $3/2$-L\'evy distribution
with the scale factor
\be 
C_0 = \frac{(G^2_{11})^2}{2\sqrt{D_1|v_1-v_2|}} = \frac{1}{\sqrt{2}} 
\ee 

On the other hand, in \cite{Bern16} the
dynamical structure function was proved rigorously to be a fundamental solution of the fractional pde
\be 
\partial_t u(x,t) = - \frac{1}{\sqrt{2}}\left[(-\Delta)^{3/4} - \nabla (-\Delta)^{1/4}\right]u(x,t).
\ee 
A Fourier transformation reproduces the mode coupling
prediction \eref{S2}, thus proving that the mode coupling
approximation is exact for this model and yields even the 
correct scale factor $C_0$ \cite{Spohnfootnote}.

\section{ Inverse relationship between the mode coupling matrices and the Hessians}
\label{app:HviaG}

The inverse relationship between the mode coupling matrices and the Hessians is
\be
H^\gamma = 2 \sum_{\lambda} (R^{-1})_{\gamma \lambda}
R^T G^\lambda R .\nonumber
\ee
Decomposing the mode coupling matrices into their two diagonal components
$G^\lambda_{\alpha\alpha}$
and the symmetric off-diagonal part 
$G^\lambda_{12}$ yields after straightforward computation
\bea 
H^\gamma 
& = &  f^\gamma_1 \left( \ba{cc} 1 &  u_+  \\  u_+  & u_+^2 \ea \right)  
+ \bar{f}^\gamma \left( \ba{cc} 2 u_- & 1+u_+u_- \\ 1+u_+u_- & 2 u_+ \ea \right) 
+ f^\gamma_2 \left( \ba{cc} u_-^2 &  u_-  \\  u_-  & 1 \ea \right) \nonumber \\
& = & 
u_+ f^\gamma_1 \left( \ba{cc} \frac{\delta + \theta}{2 J_{12}} &  1  \\  1  & \frac{\delta - \theta}{2 J_{21}} \ea \right) 
+ \frac{2 u_+}{J_{12}} \bar{f}^\gamma  \left( \ba{cc} - J_{21}  & \theta    \\ \theta &  J_{12}\ea \right)  
+ \frac{u_+ J_{21} }{J_{12}} f^\gamma_2 \left( \ba{cc}  \frac{\delta - \theta}{2 J_{12}}  &  - 1 \\  - 1  & \frac{\delta + \theta}{2 J_{21}} \ea \right)
\nonumber
\eea
where the amplitudes
\bea
f^\gamma_1 & = & \frac{2}{z_+^2} \sum_{\lambda} (R^{-1})_{\gamma \lambda} G^\lambda_{11} \nonumber\\
\bar{f}^\gamma & = & \frac{2}{z_+z_-}  \sum_{\lambda} (R^{-1})_{\gamma \lambda} G^\lambda_{12} \nonumber\\
f^\gamma_2 & = & \frac{2}{z_-^2} \sum_{\lambda} (R^{-1})_{\gamma \lambda} G^\lambda_{22} .\nonumber
\eea 
depend on the mode coupling coefficients.


\begin{thebibliography}{99}


\bibitem{Livio} Livio, M. (2008) The Golden Ratio: The Story of Phi, the World’s Most Astonishing Number, Broadway Books, New York.

\bibitem{1998Levitov} Hyun-Woo Lee and Leonid S. Levitov,
Universality in Phyllotaxis: a Mechanical Theory,  published in:
Symmetry in Plants, Series in Mathematical Biology and Medicine, eds. R. V. Jean, D. Barab\'e,  World Scientific Pub Co Inc, 1998,
also in https://arxiv.org/pdf/2101.02652.pdf

\bibitem{Kepler1611}
J. Kepler, De Nive sexangula (Godfrey Tampach, Frankfurt, 1611); English translation:
J. Kepler, On the Six-Cornered Snowflake  (Clarendon Press, Oxford, 1966)

\bibitem{2010Pletzer}
Belinda Pletzer, Hubert Kerschbaum, Wolfgang Klimesch,
When frequencies never synchronize: The golden mean and the resting EEG,
Brain Research,
vol. 1335,
2010,
Pages 91-102,
https://doi.org/10.1016/j.brainres.2010.03.074

\bibitem{2018Meisner} Gary B. Meisner and Rafael Araujo, The Golden Ratio: The Divine Beauty of Mathematics,  Race Point Publishing,  2018

\bibitem{Khinchin} A.Y. Khinchin,  Continued Fractions, 1949; 
(reprint: Dover Books on Mathematics, 1964)

\bibitem{1937Landau} L. D.  Landau,  On the theory of phase transitions,
Zh. Eksp. Teor. Fiz. 7, pp. 19–32 (1937)

\bibitem{Popk15a} 
V. Popkov, A. Schadschneider, J. Schmidt, and G.M. Sch\"utz,
Fibonacci family of dynamical universality classes,
Proc. Natl. Acad. Science (USA) \textbf{112} 12645--12650 (2015).

\bibitem{Krug97}
J. Krug,
%Krug, J.:
Origins of scale invariance in growth processes.
Adv. Phys. \textbf{46}, 139--282 (1997).

\bibitem{Halp15}
T. Halpin-Healy, K.A. Takeuchi,
%Halpin-Healy, T. and Takeuchi, K.A.:
A KPZ Cocktail-Shaken, not Stirred...
J. Stat. Phys. \textbf{160}, 794--814 (2015).

\bibitem{Popk15b}
V. Popkov, J. Schmidt, and G.M. Sch\"utz,
Universality classes in two-component driven diffusive systems.
J. Stat. Phys. \textbf{160}, 835--860 (2015).

\bibitem{Spoh15}
H. Spohn and G. Stoltz,
Nonlinear fluctuating hydrodynamics in one dimension: The case of 
two conserved fields,
J. Stat. Phys. \textbf{160}, 861--884 (2015).

\bibitem{Spoh91}
H. Spohn,
{\it Large Scale Dynamics of Interacting Particles.}
(Springer, Berlin, 1991)

\bibitem{Spoh99} 
Spohn, H.: 
%H. Spohn,
Bosonization, vicinal surfaces, and hydrodynamic fluctuation theory.
Phys. Rev. E \textbf{60}, 6411--6420 (1999).

\bibitem{Popk11} 
%V. Popkov and G.M. Sch\"utz, 
Popkov, V. and Sch\"utz, G.M.:
Transition Probabilities and Dynamic Structure Function in the ASEP Conditioned on Strong Flux,
J. Stat. Phys. \textbf {142}, 627--639 (2011).

\bibitem{Kare17}
Karevski, D., Sch\"utz, G.M.:
Conformal invariance in driven diffusive systems at high currents.
Phys. Rev. Lett. \textbf{118}, 030601 (2017)

\bibitem{Helb99} D.~Helbing, D.~Mukamel, and G.~M.~Sch\"utz,
Phys. Rev. Lett. \textbf{82}, 10--13 (1999).

\bibitem{Ross00}
M. Rossi, R. Pastor-Satorras, and A. Vespignani. Universality class of absorbing phase transitions with a conserved
field. Phys. Rev. Lett. \textbf{85} 1803--1806 (2000).

\bibitem{Gabe10}
A. Gabel, P. L. Krapivsky, and S. Redner. Facilitated asymmetric exclusion. Phys. Rev. Lett. \textbf{105} 210603 (2010).

\bibitem{Blon20}
O. Blondel, C. Erignoux, M. Sasada, and M. Simon. 
Hydrodynamic limit for a facilitated exclusion process. 
Annales de l’Institut Henri Poincar\'e, Probabilit\'es et Statistiques  \textbf{56}, 667--714 (2020).

\bibitem{Erig23}
C. Erignoux and L. Zhao,
Stationary fluctuations for the facilitated exclusion process,
arXiv:2305.13853 [math.PR] (2023)


\bibitem{Rama02}
Ramaswamy, S., Barma, M., Das, D., Basu, A.:
Phase Diagram of a Two-Species Lattice Model with a Linear Instability.
Phase Transit. \textbf{75}, 363--375 (2002)

\bibitem{Kafr03}
Kafri, Y., Levine, E., Mukamel, D., Sch\"utz, G.M., Willmann R.D.:
Phase-separation transition in one-dimensional driven models.
Phys. Rev. E \textbf{68}, 035101(R) (2003)

\bibitem{Chak16}
Chakraborty, S., Pal, S., Chatterjee, S., Barma, M.:
Large compact clusters and fast dynamics in coupled nonequilibrium systems.
Phys. Rev. E \textbf{93}, 050102(R) (2016)

\bibitem{Kipn99}
Kipnis, C., Landim, C.:
Scaling limits of interacting particle systems.
% in: {\it Grundlehren der mathematischen Wissenschaften} Vol. 320 
Springer, Berlin (1999).

\bibitem{Spoh14}
H. Spohn,
Nonlinear Fluctuating hydrodynamics for anharmonic chains.
J. Stat. Phys. {\bf 154}, 1191--1227 (2014).

\bibitem{Spoh16}
Spohn, H.:
Fluctuating Hydrodynamics Approach to Equilibrium Time Correlations for 
Anharmonic Chains. In: S. Lepri (ed.), Thermal Transport in Low Dimensions. 
From Statistical Physics to Nanoscale Heat Transfer, 
Lecture Notes in Physics 921, 107--158  (Springer, Switzerland, 2016)

\bibitem{Popk16} 
V. Popkov, A. Schadschneider, J. Schmidt, G.M. Sch\"utz,
Exact scaling solution of the mode coupling equations for 
non-linear fluctuating hydrodynamics in one dimension,
J. Stat. Mech. 093211 (2016).

\bibitem{Toth03}
B. T\'oth and B. Valk\'o,
%Onsager Relations and Eulerian Hydrodynamic Limit for Systems with Several Conservation Laws,
J. Stat. Phys. \textbf{112}, 497--521 (2003).

\bibitem{Gris11}
R. Grisi and G.M. Sch\"utz,
Current symmetries for particle systems with several conservation laws,
J. Stat. Phys. \textbf{145}, 1499--1512 (2011).

\bibitem{Cast16}
Castro-Alvaredo, O. A., Doyon, B., Yoshimura,  T.:
Emergent Hydrodynamics in Integrable Quantum Systems Out of Equilibrium.
Phys. Rev. X \textbf{6}, 041065 (2016) 
%%\doi{10.1103/PhysRevX.6.041065}
%
\bibitem{Kare19}
D. Karevski and G. M. Sch\"utz,
Charge-current correlation equalities for quantum systems far from equilibrium.
SciPost Phys. \textbf{6}, 068 (2019)

\bibitem{Doyo17}
B. Doyon and H. Spohn,
Drude Weight for the Lieb-Liniger Bose Gas,
SciPost Phys. \textbf{3}, 039 (2017).

\bibitem{Alba21}
V. Alba, B. Bertini, M., L. Piroli, P. Ruggiero,
Generalized-hydrodynamic approach to inhomogeneous quenches: correlations, entanglement and quantum effects,
J. Stat. Mech. (2021) 114004

\bibitem{DeNa22}
J. De Nardis, B. Doyon, M. Medenjak, M. Panfil,
Correlation functions and transport coefficients in generalised hydrodynamics,
J. Stat. Mech. (2022) 014002

%\bibitem{Bert16}
%B. Bertini, M. Collura, J. De Nardis, and M. Fagotti, 
%Transport in Out-of-Equilibrium XXZ Chains: Exact Profiles of Charges and Currents, 
%Phys. Rev. Lett. \textbf{117}, 207201 (2016).

\bibitem{Blyt07} 
Blythe, R.A., Evans, M.R.:
Nonequilibrium steady states of matrix-product form: a solver's guide.
J. Phys. A: Math. Theor. {\bf 40} R333--R441 (2007)

\bibitem{Cant22}
L. Cantini and A. Zahra, 
Hydrodynamic behavior of the two-TASEP,
J. Phys. A: Math. Theor. \textbf{55} 305201 (2022)

\bibitem{Gros03}
Grosskinsky S, Sch\"utz GM, Spohn H,
Condensation in the zero range process: Stationary and dynamical properties
J.~Stat.~Phys.\ {\bf  113} (3-4): 389-410 (2003).

\bibitem{Popk04}
V. Popkov and M. Salerno,
Hydrodynamic limit of multichain driven diffusive models,
Phys. Rev. E \textbf{69}, 046103 (2004).

\bibitem{Schm21}
Schmidt, J., Sch\"utz, G.M., and van Beijeren, H.: 
A lattice Gas Model for Generic One-Dimensional Hamiltonian Systems. 
J. Stat. Phys. \textbf{183}, 8 (2021). 

\bibitem{Katz84} 
S. Katz, J. L. Lebowitz, and H. Spohn, 
Nonequilibrium steady states of stochastic lattice gas models of fast ionic conductors
J. Stat. Phys. \textbf{34}, 497--537 (1984)

\bibitem{KLS}
G.M. Sch\"utz, 
Diffusion-annihilation in the presence of a driving field.
J. Phys. A: Math. Gen. \textbf{28},  3405--3415 (1995).

\bibitem{Hage01} 
J.S. Hager, J. Krug, V. Popkov and G.M. Sch\"utz, 
Phys. Rev. E \textbf{63}, 056110 (2001).

\bibitem{Dier13}
M. Dierl, M. Einax, P. Maass,
One-dimensional transport of interacting particles: Currents, density profiles, phase diagrams and symmetries
Phys. Rev. E \textbf{87}, 062126 (2013).


\bibitem{Derr98}
%Derrida, B.: 
B. Derrida,
An exactly soluble nonequilibrium system: The asymmetric simple exclusion process.
Phys. Rep. \textbf{301}, 65--83 (1998).

\bibitem{Ligg99} 
Liggett, T.M.:
Stochastic Interacting Systems: Contact, Voter and Exclusion Processes.
Springer, Berlin (1999)

\bibitem{Schu01} 
Sch\"utz, G.M.:
Exactly solvable models for many-body systems far from equilibrium, in: 
Phase Transitions and Critical Phenomena Vol. 19, 
C. Domb and J. Lebowitz (eds.), 1--251,
Academic Press, London (2001).

\bibitem{Guio99}
H. Guiol, 
Some Properties of k-Step Exclusion Processes,
J. Stat. Phys. \textbf{94}, 495--511 (1999).

\bibitem{Guio02}
%Freire, M.V., Guiol, H., Ravishankar, K., Saada, E.:
M.V. Freire, H. Guiol, K. Ravishankar, and E. Saada, 
Microscopic structure of the k-step exclusion process,
Bull. Braz. Math. Soc. \textbf{33}, 319--339 (2002).






%\bibitem{Schu18}
%G.M. Sch\"utz, 
%%Sch\"utz, G.M.:
%On the Fibonacci universality classes in nonlinear fluctuating hydrodynamics,
%In: P. Gon{\c c}alves and A.J. Soares (eds.)
%From Particle Systems To Partial Differential Equations V, 
%%Springer Proceedings in Mathematics & Statistics,
%pp. 149--167, Springer, New York (2018)

\bibitem{Bern09}
C. Bernardin and S. Olla, 
Non-equilibrium macroscopic dynamics of chains of
anharmonic oscillators,
http://cermics.enpc.fr/~stoltz/Olla/lecture\_1.pdf
(2009)

%
\bibitem{Bern12}
C. Bernardin and G. Stoltz, 
Anomalous diffusion for a class of systems with two conserved quantities,
Nonlinearity \textbf{25}, 1099--1133 (2012).

\bibitem{Bern16}
C. Bernardin, P. Gon{\c c}alves, and M. Jara, 
3/4-fractional superdiffusion in a system of harmonic oscillators perturbed by a conservative noise,
Arch. Rational Mech. Anal. \textbf{220}, 505--542 (2016).


\bibitem{Spohnfootnote}
After completing this computation, P. Gon{\c c}alves 
mentioned to us that the correctness of the scale factor $C_0$ was communicated to her by H. Spohn some years ago. Since we are not aware of a published version of this observation we decided to keep it in this manuscript, without claiming originality.

\end{thebibliography}
\end{document}